\documentclass{article}

\usepackage{arxiv}

\usepackage[utf8]{inputenc} 
\usepackage[T1]{fontenc}    
\usepackage{hyperref}       
\usepackage{url}            
\usepackage{booktabs}       
\usepackage{amsfonts}       
\usepackage{nicefrac}       
\usepackage{microtype}      
\usepackage{lipsum}
\usepackage{fancyhdr}       
\usepackage{graphicx}       
\graphicspath{{Figures/}}     

\title{COVID-Net USPro: An Open-Source Explainable Few-Shot Deep Prototypical Network to Monitor and Detect COVID-19 Infection from Point-of-Care Ultrasound Images}

\author{
  Jessy Song \\
  Department of Systems Design Engineering \\
  University of Waterloo \\
  Waterloo, ON N2L 3G1, Canada\\
 \And
  Ashkan Ebadi \\
  Digital Technologies Research Centre \\
  National Research Council Canada \\
  Toronto, ON M5T 3J1, Canada\\
  \texttt{ashkan.ebadi@nrc-cnrc.gc.ca} \\
  \And
  Adrian Florea \\
  Department of Emergency Medicine \\
  McGill University \\
  Montreal, QC H4A 3J1, Canada \\
  \And
  Pengcheng Xi, Stéphane Tremblay \\
  Digital Technologies Research Centre \\
  National Research Council Canada \\
  Ottawa, ON K1A 0R6, Canada \\
  \And
  Alexander Wong \\
  Department of Systems Design Engineering \\
  University of Waterloo \\
  Waterloo, ON N2L 3G1, Canada \\
}

\begin{document}
\maketitle

\begin{abstract}
As the Coronavirus Disease 2019 (COVID-19) continues to impact many aspects of life and the global healthcare systems, the adoption of rapid and effective screening methods to prevent further spread of the virus and lessen the burden on healthcare providers is a necessity. As a cheap and widely accessible medical image modality, point-of-care ultrasound (POCUS) imaging allows radiologists to identify symptoms and assess severity through visual inspection of the chest ultrasound images. Combined with the recent advancements in computer science, applications of deep learning techniques in medical image analysis have shown promising results, demonstrating that artificial intelligence-based solutions can accelerate the diagnosis of COVID-19 and lower the burden on healthcare professionals. However, the lack of a huge amount of well-annotated data poses a challenge in building effective deep neural networks in the case of novel diseases and pandemics. Motivated by this, we present COVID-Net USPro, an explainable few-shot deep prototypical network, that monitors and detects COVID-19 positive cases with high precision and recall from minimal ultrasound images. COVID-Net USPro achieves 99.65\% overall accuracy, 99.7\% recall and 99.67\% precision for COVID-19 positive cases when trained with only 5 shots. The analytic pipeline and results were verified by our contributing clinician with extensive experience in POCUS interpretation, ensuring that the network makes decisions based on actual patterns.
\end{abstract}


\section{Introduction}
The Coronavirus Disease 2019, or COVID-19, caused by severe acute respiratory syndrome coronavirus 2 (SARS-CoV-2), has been continuously impacting individual’s well-being and the global healthcare systems \cite{covidxus2022}. Despite the vaccination efforts, policies and regulations in place, due to the rapid transmission of the virus and waves of rising cases, the development of effective screening and risk stratification methods remains to be a critical need in controlling the disease \cite{NCBICOVIDTreatment}. Various types of diagnostic tools, including reverse transcription-polymerase chain reaction (RT-PCR), rapid antigen detection tests, and antibody tests, have been developed and adapted globally to increase the rate of screening. While RT-PCR has been the gold standard test for diagnosing COVID-19, the technique involves large labour and laboratory resources and is time-consuming \cite{RapidTestsCOVID}. Other rapid antigen tests and antibody tests with varying sensitivity are also less reliable in comparison to RT-PCR tests \cite{RapidTestsCOVID}. 
 
For people with significant respiratory symptoms, medical imaging is used to identity the disease and assess the severity of the disease progression \cite{DetectCOVIDDL}. Under this protocol, a computed tomography (CT) scan, chest X-ray (CXR), or point-of-care ultrasound (POCUS) imaging can be performed and used clinically as an alternative diagnostic tool \cite{NCBICOVIDTreatment}. To make a diagnosis, acute care physicians and radiologists visually inspect the radiographic images to find patterns related to symptoms and to assess severity of COVID-19 infection and deformation \cite{RapidTestsCOVID}. During times of high transmission rate of COVID-19, large influx of patients increases the burden on clinicians and radiologists. Medical image processing and artificial intelligence (AI) can assist in reducing this burden and accelerate the diagnostic and decision-making process, as existing models and algorithms continue to improve and the amount of available medical image data continues to grow \cite{COVIDNet,COVIDNetCTGunraj2020Ver2,COVIDNetUSMacLean2021}. 

Different imaging modalities, including CT scan, X-ray, and ultrasound may be used in the diagnosis of COVID-19 and offer varying diagnostic values \cite{MedicalImageTechniques}. Chest CT scan is the most sensitive imaging modality in the initial diagnosis and management of confirmed cases, but it is more expensive and time-consuming \cite{MedicalImageTechniques, COVIDNet}. In contrast, ultrasound imaging is more accessible and portable, cheap, and safer as radiation is not involved during the examination, which are desirable properties for its usage \cite{MedicalImageTechniques}, especially in resource-limited settings/environments/areas/regions.

Deep learning usually requires a large set of training examples \cite{COVIDNetCTGunraj2020,COVIDNetUSMacLean2021,DetectCOVIDDL}. However, due to the nature of novel diseases, the availability of such a huge amount of well-annotated data poses a great challenge to the learning algorithms. Few-shot learning is an approach where model is trained to classify new data based on a limited number of samples exposed in training \cite{protonetSnell}. This resembles how humans learn, as we can recognize new object classes from very few instances, different from other current machine learning techniques that require large amount of data to achieve similar performance \cite{protonetSnell}. Since the few-shot model requires less data to train, the computational costs in the process is also significantly reduced \cite{protonetSnell}. These properties make it an appropriate and promising approach for COVID-19 or rare disease diagnosis. One approach for few-shot learning is metric-based learning. As a few-shot metric-based learning approach, prototypical networks (PN) perform classification by computing distances to prototype representations of each class \cite{protonetSnell}. PN has shown state-of-the-art (SOTA) results on other datasets/domains (e.g., \cite{PNSegmentation, PNTextClassification, PNHistopathology}), proving that some simple design decisions can yield significant improvements over other complicated architectures and meta-learning approaches \cite{protonetSnell}. 

Motivated by the needs for fast and effective alternative screening solutions and considering ultrasound imaging advantages, we present an open-source explainable deep prototypical network, called COVID-Net USPro, that learns to detect COVID-19 positive cases with high precision and recall from a very limited number of lung ultrasound (LUS) images. When trained with only 5 shots, COVID-Net USPro classifies between positive and negative COVID-19 cases with 99.65\% overall accuracy, 99.7\% recall and 99.67\% precision for COVID-19 positive cases. Intensive experimentation was conducted (e.g.,  testing different image encoders, varying training conditions and number of classes to optimize the network) to assess the performance of COVID-Net USPro network. To ensure the network’s fairness and accountability, network benefits from an explainability module, assessing decisions with visual explanation tools, i.e., Grad-CAM \cite{GradCam} and GSInquire \cite{gsinquire}. Moreover, our contributing clinician (A.F.) carefully verified and validated the pipeline and produced results to ensure the validity of the proposed solution from the clinical perspective.

\subsection{Related Work}\label{sec:related_work}
There are several studies that aim to apply deep learning into the screening and detection of COVID-19 positive cases. As an open-source and open-access initiative, the COVID-Net \cite{COVIDNetInitiative, COVIDNet, COVIDNetCTGunraj2020, COVIDNetUSMacLean2021} includes research on the application of deep learning neural networks using multitude of image modalities, such as CT, X-ray, and ultrasound images. Multiple works have demonstrated the effectiveness of deep learning in the classification of CT and X-ray images. For example, COVID-Net CXR \cite{COVIDNetCXR3}, a tailored deep convolutional neural network (DCNN/CNN) for detection of COVID-19 cases from chest X-ray images, has achieved an overall accuracy of 98.3\% and 97.5\% sensitivity for COVID-19 cases. Another work by Ozturk et al. proposed a DCNN based on the DarkNet model used for the you only look once (YOLO) real time object detection system to classify X-ray images, which achieves 98.08\% accuracy for binary COVID-19 cases detection \cite{COVIDYOLODCNNXray}. Research by Afshar et al. proposed a capsule CNN-based network called COVID-CAPS \cite{COVIDCAPS} which achieved over 98\% accuracy and specificity using a limited amount of X-ray images. COVID-Net CT \cite{COVIDNetCTGunraj2020Ver2}, a deep neural network for detection of COVID-19 from CT images, scored 96.2\% in sensitivity and 99\% in specificity for COVID-19 cases. Potential of including both CT-scan and X-ray images for classification is also explored, with research by Thakur and Kumar demonstrating a DCNN-based model achieving over 99\% accuracy and precision for COVID-19 detection using images of both modalities \cite{XrayCTCOVIDCNN}. For ultrasound images, custom neural network such as COVID-Net US \cite{COVIDNetUSMacLean2021} was constructed and tailored to COVID-19 case detection. The network achieved an area under receiver operating curve (AUC) of over 98\% when trained with positive COVID-19 and negative normal case images. Research by Diaz-Escobar et al. \cite{COVIDUSDL} also leveraged pre-trained neural networks such as VGG19 \cite{VGG}, InceptionV3 \cite{InceptionV3}, and ResNet50 \cite{ResNet} in the detection of COVID-19 using ultrasound images and achieved 89.1\% accuracy and AUC of 97.1\%. One limitation of using a custom deep neural network in most of the existing research is the need for a large amount of training data, where in mentioned works above, datasets all surpassed 10,000 total images \cite{COVIDNetUSMacLean2021,COVIDNetCTGunraj2020}. 

Application of few-shot learning techniques has also been investigated. For example, MetaCOVID, proposed by Shorfuzzaman et al \cite{MetaCOVID}, is a Siamese neural network framework with contrastive loss for few-shot diagnosis of COVID-19 infection using CXR images. The performance of the best network achieved an accuracy of 95.6\% and AUC of 97\% when trained under a 3-way, and tested in a 10-shot setting \cite{MetaCOVID}. In \cite{covidnet_fewse_2021}, a deep siamese convolutional network, called COVID-Net FewSE, is able to detect COVID-19 positive cases with 90\% recall and accuracy of 99.7\% when the network is provided with only 50 observations in the training phase. In the work by Karnes et al. \cite{AdaptiveFewShotCOVIDUS}, the possibility of using adaptive few-shot learning for ultrasound COVID-19 detection is examined, and the increasing performance with the increasing number of shots is investigated. Although the feasibility of adopting few-shot learning techniques for COVID-19 detection from medical imaging has been already investigated, analysis on network’s interpretability is either missing or inadequate and lacks clinician validation, which limits the full understanding of the network and whether data interpretation process aligns with real clinical settings. 

Our contribution is at least three folds: 1) We presents a high-performing network (99.65\% accuracy) trained with only 5 shots, while other works achieving similar performance require larger numbers of training examples, 2) COVID-Net USPro is an explainable network, as demonstrated by analysis from two explainability visualization tools and clinician validation, and 3) COVID-Net USPro is open-sourced and available to the public, which helps promote reproducibility and  accessibility of AI in healthcare and encourage further innovation. 

The remainder of this paper is as follows. Section \ref{sec:mat_meth} explains data, techniques, and the experiments conducted to assess the network performance in details. Section \ref{sec:results} presents findings from the analysis. Findings are then discussed in Section \ref{sec:discussion} where some limitations of the research and future directions are also presented.

\section{Data and Methodology}\label{sec:mat_meth}

\subsection{Data}
The \href{https://github.com/nrc-cnrc/COVID-US}{COVIDx-US} dataset v1.4. \cite{covidxus2022} is used for this study. COVIDx-US is an open-access benchmark dataset of lung ultrasound imaging data that contains 242 videos and 29,651 processed images of patients with COVID-19 infection, non-COVID-19 infection, other lung conditions, and normal control cases. The dataset provides LUS images captured with two kinds of probe, linear probe which produces a square or rectangular image, or convex probe, which allows for a wider field of view \cite{USProbes}. Due to the difference in field of view and low numbers of COVID-19 positive examples with linear probe, combining the linear and convex probe data in training may increase noise and influence the performance of the network and hence, linear probe data are excluded in this study. A total number of 25,262 convex LUS images are then randomly split into train set containing 90\% of images in each class and test set with the remaining 10\% of images, ensuring all frames from each video are either in train or test set to avoid data leakage. All images are rescaled to $224\times224$ pixels to keep the images across entire dataset consistent. The dataset is further augmented by rotating each image by 90°, 180°, 270°, resulting in a total of 101,048 images ($25262\times4$). This rotation technique is an appropriate method for increasing the dataset size, as it keeps the images and areas of interest for clinical decisions unaltered and in-bound \cite{dataAugmentation}.

\subsection{Methodology}
\begin{figure}
  \centering
  \includegraphics[width=13.5 cm]{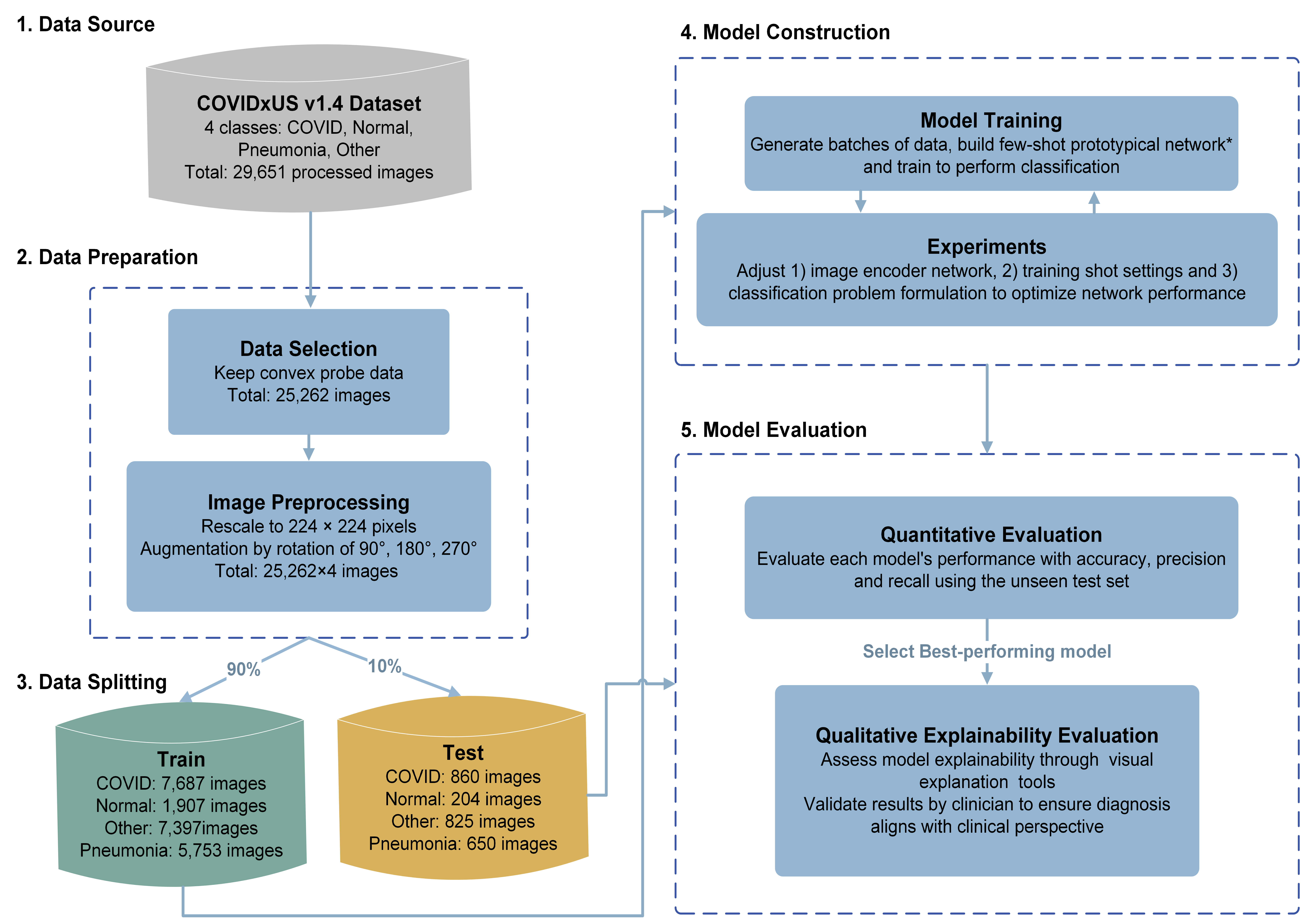}
  \caption{High-level conceptual flow of the Analysis.}
  \label{fig:overview}
\end{figure}

COVID-Net USPro is a prototypical few-shot learning network that trains in an episodic learning setting, using a distance metric for assessing similarities between a set of unlabelled data, i.e., query set, and labelled data, i.e., support set. Labelled data can be used to compute a single \textit{prototype} representation of the class, and unlabelled data are assigned to the class of the prototype they are closest to. A prototypical network \cite{protonetSnell} is based on this idea that there exists an embedding in which points in a class cluster around a single prototype representation for the class. During the training phase, a neural network is used to learn the non-linear mapping of the inputs to an embedding space, and a class prototype is computed as the mean of its support set data in the embedding space. Classification is then done by finding the nearest class prototype for each query point based on a specified distance metric. An episodic approach is used to train the model, where in each training episode, the few-shot task is simulated by sampling the data point in mini-batches to make the training process consistent with the testing environment. Performance of the network is evaluated using the test dataset, and both quantitative analysis based on accuracy, precision and recall and qualitative explainability analysis are conducted. An high-level conceptual flow of the Analysis is presented in Figure~\ref{fig:overview}. 

We defined the classification problem as a $K$-way $N$-shot episodic task, where $K$ denotes the number of classes present in the dataset and $N$ denotes the number of available image examples for each class in each episode. For a given dataset, $N$ images from each of the $K$ classes are sampled to form the support set, and another $M$ images from each class are sampled to form the query set. The network then aims to classify the images of the query set based on the $K*N$ total images presented in the support set. In this work, we formulated the problem as a 2-way, 3-way and 4-way classification problem. Details are included under section \ref{sec:experimentsProbType}. 

The few-shot classification with prototypical network can be summarized into three steps: 1) encoding of the images, 2) generating class prototypes, and 3) assigning labels to query samples based on distance to the class prototypes. Let's $S=\{(\textbf{x}_{(1,s)},{y}_{(1,s)}),\ldots,(\textbf{x}_{(N,s)}, {y}_{(N,s)})\}$ and  $Q=\{(\textbf{x}_{(1,q)},y_{(1,q)}), \ldots,(\textbf{x}_{(N,q)}, y_{(N,q)})\}$ be the support and query sets respectively, where each $x_i \in R^D$ is a $D$-dimensional example feature vector and $y_i \in \{1,\ldots K\}$ is the label of the example. The prototypical network embodies an image encoder $f_\phi: R^D \rightarrow R^H$ that transforms each image $x_i$ onto a $H$-dimensional embedding space where images of the same class cluster together. Class prototypes are then generated for each class by averaging the embedding image vectors in the support set, where $v_k=\frac{1}{N}\sum_{i=1}^{N}{f_\phi({{x}_{i,s}}^{(k)})}$ denotes the prototype of class $k$ \cite{protonetSnell}. To classify the query image, a distance metric is used where distances between the embedding vector of a query image and each of the class prototypes are computed. In this work, squared Euclidean distance $d\left(v,q\right)=\left||v-q|\right|=\sqrt{\sum\left(v_i-q\right)^2}$ is used, where $q$ is the embedding vector of the query image and $v_i$ is the embedding vector of the $i$-th prototype. After distances are computed, a SoftMax function is applied over distances to the prototypes to compute the probabilities of the query image being in each class. The class with the highest probability is then assigned to the query image. 

\begin{figure}
\centering
\includegraphics[width=13.5 cm]{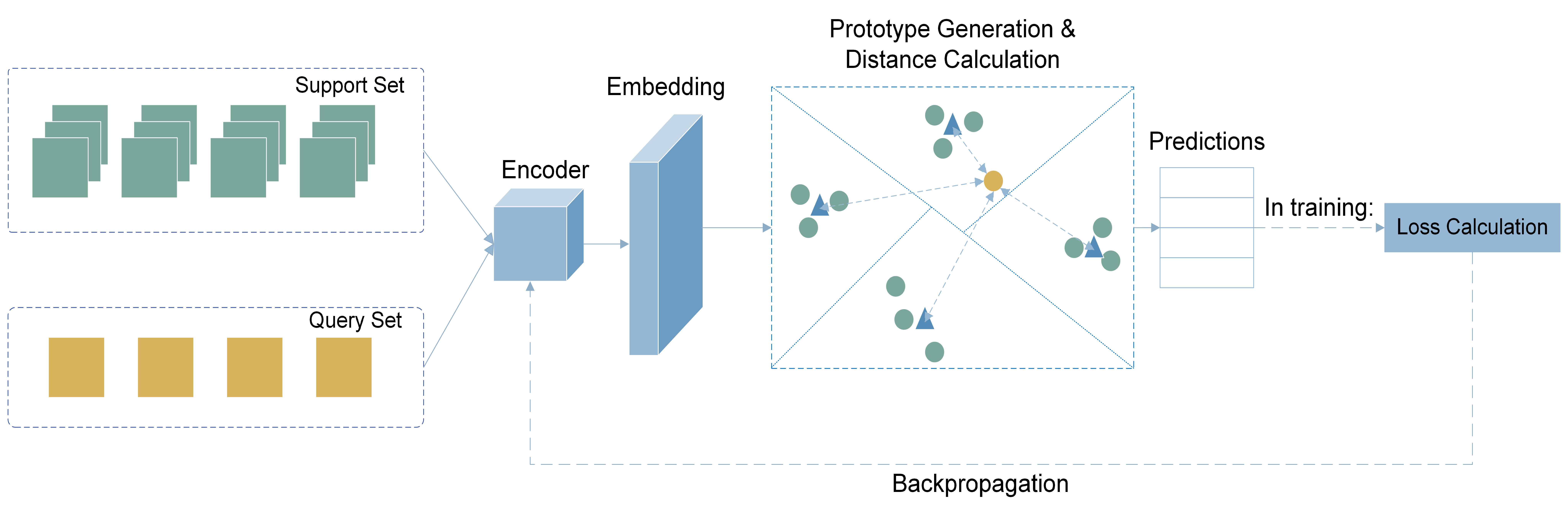}
\caption{COVID-Net USPro, network architecture design.\label{fig:architectureFlow}}
\end{figure}

In the training phase, the network learns by minimizing a loss function, i.e., the negative log-SoftMax function ($J=-\log{\left(p\left(y=k\middle| x\right)\right)}$) of the true class $k$ via an optimizer for which we use an Adam optimizer with an initial learning rate of 0.001, and reduced if loss is not improved after 3 epochs. In each episode, a subset of data points is randomly selected, forming support and query set. Loss term is calculated at the end of each training episode. To facilitate effective training process and prevent over-fitting, early stopping is implemented to stop the training process when loss term is not improved after 5 epochs. A total of 10 epochs is set for all training processes and 200 episodes is set for each training epoch. Figure~\ref{fig:architectureFlow} presents an architecture design overview of the COVID-Net USPro network. 

Trained model's performance is evaluated quantitatively and qualitatively. In quantitative analysis, model's accuracy, precision and recall for each class are reported. In qualitative analysis, model explainability is investigated and visualized. Explainable Artificial Intelligence (XAI) has been an important criterion when assessing whether neural networks can be applied to real clinical settings \cite{AIExplainability}. While AI-driven systems may show high accuracy and precision in analyzing medical images, lack of reasonable explainability will spark criticism to the network’s adoption \cite{AIExplainability}. COVID-Net USPro’s explainability is assessed using two approached, i.e., Gradient-weighted Class Activation Map (Grad-CAM) \cite{GradCam} and GSInquire \cite{gsinquire}, on a selected dataset containing correctly classified COVID-19 and normal cases with high confidence (i.e., $>99.9\%$ probability) as well as falsely predicted COVID-19 and normal cases. Grad-CAM generates a visual explanation of the input image using the gradient information flowing into the last convolutional layer of the convolutional neural network (CNN) encoder and assigns importance values to each neuron for making a classification decision \cite{GradCam}. The output is a heatmap-overlayed image that shows the regions that impact the particular classification decision made by the network \cite{GradCam}. The other tool GSInquire identifies the critical factors in an input image that are shown to be integral to the decisions made by the network in a generative synthesis approach \cite{gsinquire}. The result is an annotated image highlighting the critical region, which drastically changes the classification result if removed \cite{gsinquire}. Results from both tools are reviewed by a clinician with experience in analysis of ultrasound images to assess whether clinically important patterns are captured by the network.

\subsection{Experiment Settings}\label{sec:experiments}
We comprehensively assess the performance of COVID-Net USPro in detecting COVID-19 positive cases from ultrasound images by testing various training conditions such as image encoders, number of shots available for training, and classification task types. Details are further discussed in this section. 

\subsubsection{Image Encoders}
To leverage the power of transfer learning, multiple encoders are experimented, including but not limited to the ResNet and VGG-based models \cite{ResNet, VGG}. Pre-trained models refer to using model parameters pre-trained on ImageNet \cite{ImageNet}. Here, we report 4 best encoders with respect to our research objectives:

\begin{itemize}
\item \textbf{ResNet18L1:} Pre-trained ResNet18 \cite{ResNet}, with trainable parameters on the final connected layer and setting out features as the number of classes. Model 1 is regarded as the baseline model for encoders, as it contains the least number of layers and retrained parameters.
\item \textbf{ResNet18L5:} Pre-trained ResNet18 \cite{ResNet}, with trainable parameters on the last 4 convolutional layers and final connected layer. Out features set to the number of classes. 
\item \textbf{ResNet50L1:} Pre-trained ResNet50 \cite{ResNet}, with trainable parameters on the final connected layer and setting out features as the number of classes. 
\item \textbf{ResNet50L4:} Pre-trained ResNet50 \cite{ResNet}, with trainable parameters on the last 3 convolutional layers and final connected layer. Out features set to the number of classes. 
\end{itemize}

\subsubsection{Number of Training Shots}
The optimal number of shots for maximized performance is tested by training models under 5, 10, 20, 30, 40, 50, 75, and 100-shot scenarios. For selected models showing steady increase of performance over increasing shots, 150 and 200-shot conditions are tested to verify that the maximum performance is reached at 100-shot. To ensure training process is faithful to the testing environment, the number of example shots for each class presented in each episode is the same in support and query set in both training and testing. For example, in 5-shot scenario, 5 images in each class are presented for both support set and query set in training, and the same follows in testing. 

\subsubsection{Problem Formulation}
\label{sec:experimentsProbType}
As the ability of the model to correctly identify COVID-19 positive cases is valued the most in comparison to other classes, the classification problem for identifying COVID-19 was formulated in 3 different scenarios as follows, in an ascending order of data complexity: 

\begin{itemize}
\item \textbf{2-way classification:} Data from all 3 other classes, namely 'normal' class, 'non-COVID-19' class and 'other' class, are viewed as a combined COVID-19 negative class. The network learns from COVID-19 positive and COVID-19 negative dataset in this setting. 
\item \textbf{3-way classification:} As the 'other' class contains data from multiple different lung conditions, it has the highest variations and may disrupt network’s learning process due to the lack of uniformity in the data compared with COVID-19, normal or non-COVID-19 class. In 3-class classification, the ‘other’ class is excluded, and the network is trained to classify the remaining three classes. 
\item \textbf{4-way classification:} As the dataset contains four classes, the four-class classification condition remains this setting and network is trained to classify 'COVID-19', 'normal', 'non-COVID-19' and 'other' class. 
\end{itemize}

\section{Results}\label{sec:results}
This section summarizes the quantitative performance results of all combination of experiment settings listed in Section \ref{sec:experiments} as well as the results of the network explainability analysis. 

\subsection{Quantitative Performance Analysis}
The performance of COVID-Net USPro is evaluated using the overall accuracy, and the precision and recall for each class. As the performance of the model to diagnose COVID-19 positive cases is the most important for current clinical use case, precision and recall for only COVID-19 case is reported below. To reduce table size, Table~\ref{table_1} only summarizes the performance of the network under 5-shot and 100-shot scenarios for encoders that scored over 80\% across all evaluated metrics. For full performance results of all shot settings and precision, recall for all classes, please refer to project repository: \href{[http://www.anonymous]}{[www.anonymous]}. 

\begin{table}
 \caption{Summary of classification results for 5-shot and 100-shot conditions.}
  \centering
  \begin{tabular}{llllll}
    \toprule
    \textbf{Scenario}	& \textbf{No. shots}	& \textbf{Model}	& \textbf{Accuracy}	& \textbf{Precision}	& \textbf{Recall}\\
    \midrule
    2-way		& 5			& ResNet18L1			& 0.9420			& 0.9486			& 0.9460\\
    2-way		& 5			& ResNet18L5			& 0.9930			& 0.9925			& 0.9950\\
    2-way		& 5			& ResNet50L1			& 0.9525			& 0.9570			& 0.9560\\
    2-way		& 5			& ResNet50L4			& \textbf{0.9965}			& \textbf{0.9967}			& \textbf{0.9970}\\
    2-way		& 100			& ResNet18L1			& 0.9758			& 0.9764			& 0.9755\\
    2-way		& 100			& ResNet18L5			& \textbf{1.0000}			& \textbf{1.0000}			& \textbf{1.0000}\\
    2-way		& 100			& ResNet50L1			& 0.9963			& 0.9964			& 0.9962\\
    2-way		& 100			& ResNet50L4			& 0.9999			& 0.9999			& \textbf{1.0000}\\
    \hline
    3-way		& 5			& ResNet18L1			& 0.9570			& 0.9606			& 0.9510\\
    3-way		& 5			& ResNet18L5			& \textbf{0.9987}			& \textbf{0.9992}			& \textbf{0.9970}\\
    3-way		& 5			& ResNet50L1			& 0.9945			& 0.9508			& 0.9660\\
    3-way		& 5			& ResNet50L4			& 0.9947			& 0.9942			& 0.9940\\
    3-way		& 100			& ResNet18L1			& 0.9867			& 0.9833			& 0.9853\\
    3-way		& 100			& ResNet18L5			& \textbf{1.0000}			& \textbf{1.0000}			& \textbf{1.0000}\\
    3-way		& 100			& ResNet50L1			& 0.9977			& 0.9970			& 0.9975\\
    3-way		& 100			& ResNet50L4			& \textbf{1.0000}			& \textbf{1.0000}			& \textbf{1.0000}\\
    \hline
    4-way		& 5			& ResNet18L1			& 0.8627			& 0.9281			& 0.8710\\
    4-way		& 5			& ResNet18L5			& 0.9817			& \textbf{0.9975}			& \textbf{0.9970}\\
    4-way		& 5			& ResNet50L1			& 0.9392			& 0.9640			& 0.9540\\
    4-way		& 5			& ResNet50L4			& \textbf{0.9850}			& 0.9917			& 0.9930\\
    4-way		& 100			& ResNet18L1			& 0.9385			& 0.9742			& 0.9704\\
    4-way		& 100			& ResNet18L5			& 0.9884			& \textbf{1.0000}			& \textbf{1.0000}\\
    4-way		& 100			& ResNet50L1			& 0.9813			& 0.9947			& 0.9955\\
    4-way		& 100			& ResNet50L4			& \textbf{0.9902}			& \textbf{1.0000}			& \textbf{1.0000}\\
        \bottomrule
  \end{tabular}
  \label{table_1}
\end{table}

\begin{figure}
\centering
\includegraphics[width=13.5 cm]{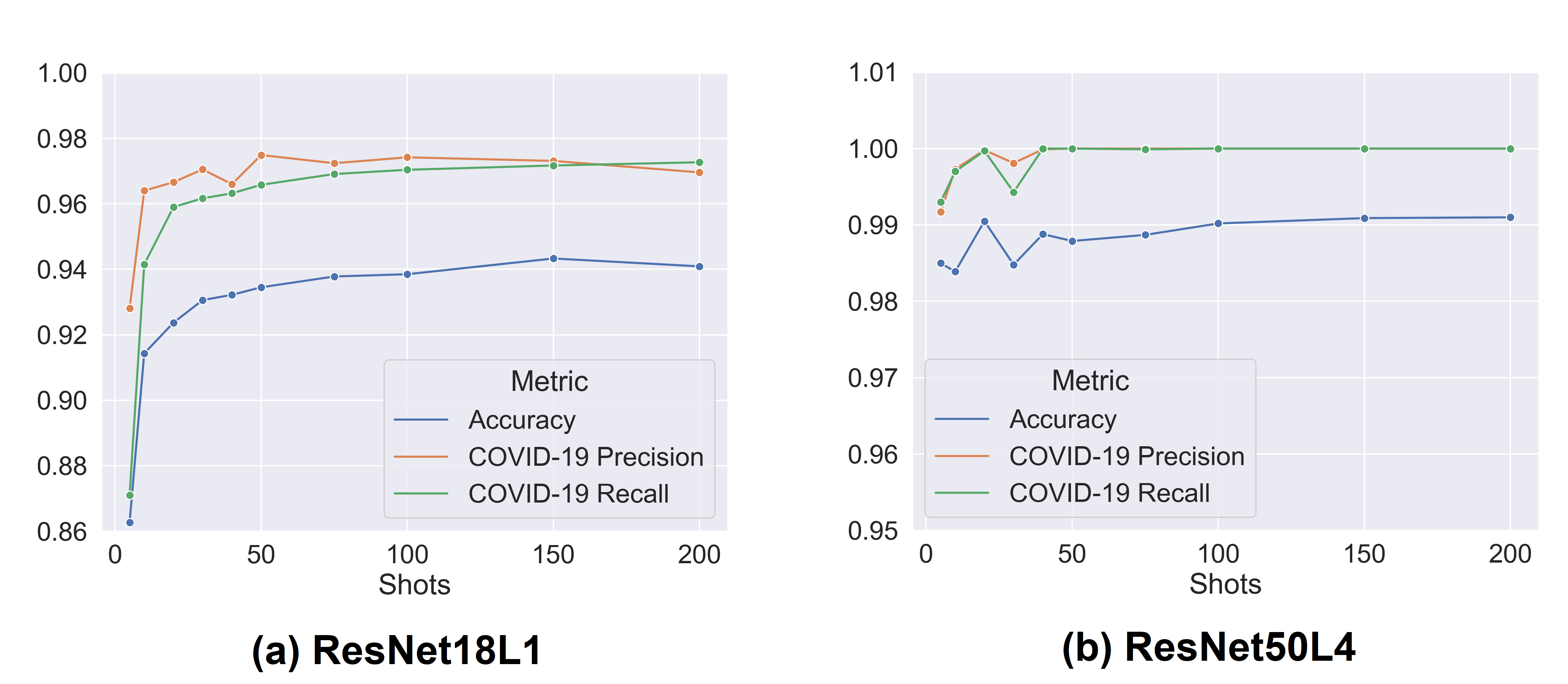}
\caption{Performance results with increasing shots trained under 4-class condition: (\textbf{a}) Pre-trained ResNet18 with trainable parameters on the final connected layer and setting out features as the number of classes (ResNet18L1). (\textbf{b})  Pre-trained ResNet50 with trainable parameters on the last 3 convolutional layers and final connected layer (ResNet50L4).\label{fig:shotsGraphs}}
\end{figure}

Across all classification types and models, performance is better under 100-shots training scenario than in 5-shot, with performance metrics increasing from 5-shot and plateauing after ~75-shot, as shown in Figure~\ref{fig:shotsGraphs}. ResNet networks demonstrate the ability to classify COVID-19 with precision and recall above 87\% consistently under both 5-shot and above 99\% under 100-shot condition. As seen in Table~\ref{table_1}, the increasing classes in 3-way and 4-way classification types reduces the performance of the network, as the classification is more complex given larger number of classes. However, this performance difference among the three classification types is reduced when the number of shots increases, as more examples available in training improves the network’s ability to distinguish between multiple classes. Among the four models, deeper models (i.e., those with ResNet50 as encoder) perform better in all classification types and shot conditions. In addition, models with re-trained final convolutional layers parameters (model ResNet18L5 and ResNet50L4) using the ultrasound images achieve higher accuracy, precision, and recall. Therefore, it can be said that while using pre-trained parameters and simpler models reduce the computational complexity and space, tailoring parameters on the final 3-4 convolutional layers to the ultrasound images and deeper image encoding boosted performance to above 99\%. 

In 2-way and 3-way classification, it is also observed that the precision and recall for classes other than COVID-19 achieve similar magnitude as the COVID-19 class. In the 4-way case, the precision and recall for ‘other’ class is around 2-3\% lower than those for ‘non-COVID-19’, ‘normal’ and ‘COVID-19’ classes. This is expected since the ‘other’ class covers various lung conditions/diseases that encompass a larger range of image features and variations. Overall, with precision and recall achieving similar magnitude for all cases in 2-way, 3-way and 4-way classification, the network also demonstrates the ability to distinguish between multiple diseases. In comparison to studies outlined in Section~\ref{sec:related_work}, the performance of COVID-Net USPro networks tailored to ultrasound images with re-trained parameters is improved. Accuracy of ResNet50L1 and ResNet50L4 exceeds 98\% under 4-way 5-shot setting, while other work such as MetaCOVID \cite{MetaCOVID}, which also applied a few-shot approach, achieved 95.6\% accuracy under 3-way, 10-shot setting. Additionally, the sensitivity of COVID-Net USPro for COVID-19 cases are also higher than networks trained with other image modality data such as X-ray or CT, where they scored 97.5\% in the best performing case \cite{COVIDNetCTGunraj2020Ver2}.

\subsection{Clinical Validation and Network Explainability Analysis}
In addition to the intensive quantitative performance analysis, we clinically validated the network output to ensure that the network captures important patterns in the ultrasound images. For this purpose, our contributing clinician (A.F.) reviewed a randomly selected set of images and reported his findings and observations. Our contributing clinician (A.F.) is an Assistant Professor in the Department of Emergency Medicine and the ultrasound co-director for undergraduate medical students at McGill University. He is practicing Emergency Medicine full-time at Saint Mary’s Hospital in Montreal. 

\begin{figure}
\centering
\includegraphics[width=13.5 cm]{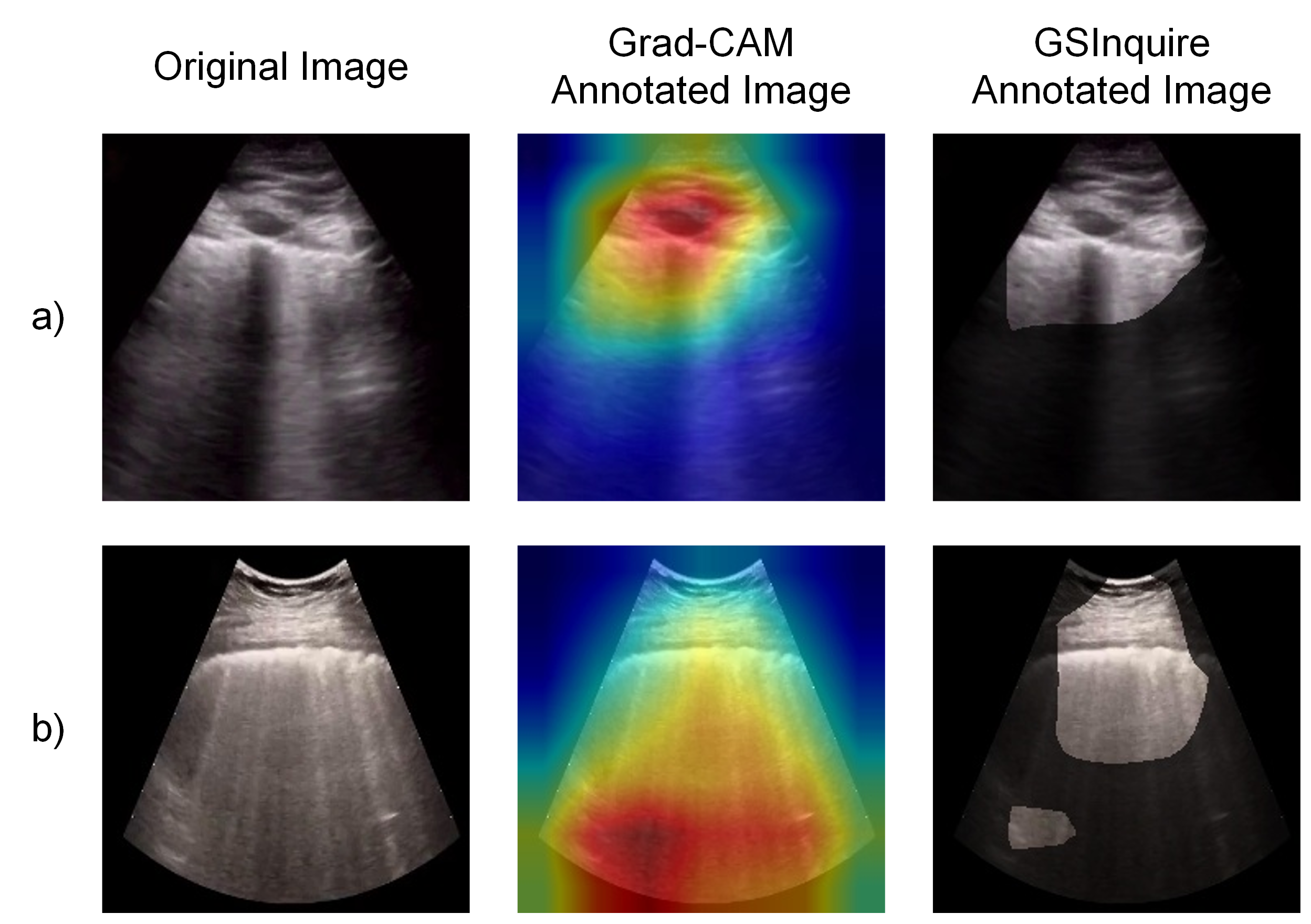}
\caption{\label{fig:true_pos}COVID-19 positive case examples correctly classified by COVID-Net USPro with high confidence:  (\textbf{a}) an example of wrong decision factors. (\textbf{b}) an example of a decision made based on the disease-related patterns.}
\end{figure}

Figure~\ref{fig:true_pos} presents two select ultrasound images of COVID-19 positive cases, annotated by Grad-CAM and GSInquire, as examples, that were reviewed. As seen, the annotated images contain the lung pleura region at the top of the image, while the second example (Figure~\ref{fig:true_pos}-b) also marks the bottom region with high importance. B-lines, or the light comet-tail artifacts extending from pleura to the bottom of the image, and the presence of dark regions interspacing the B-lines at the bottom part of the image corresponding to signs of lung consolidation are indicators of abnormality \cite{AlinesBlines}. Hence, the visual annotations for the second example (Figure~\ref{fig:true_pos}-b) are more representative for disease-related patterns within the ultrasound image. Figure~\ref{fig:true_pos}-a is one of the examples where the model considers the rib as a structure of interest, which is not the abnormality, leading to classify the images as a COVID-19 positive case. Hence, although the model correctly classified the image, the decision was made based on invalid clinical factors. 

\begin{figure}
\centering
\includegraphics[width=13.5 cm]{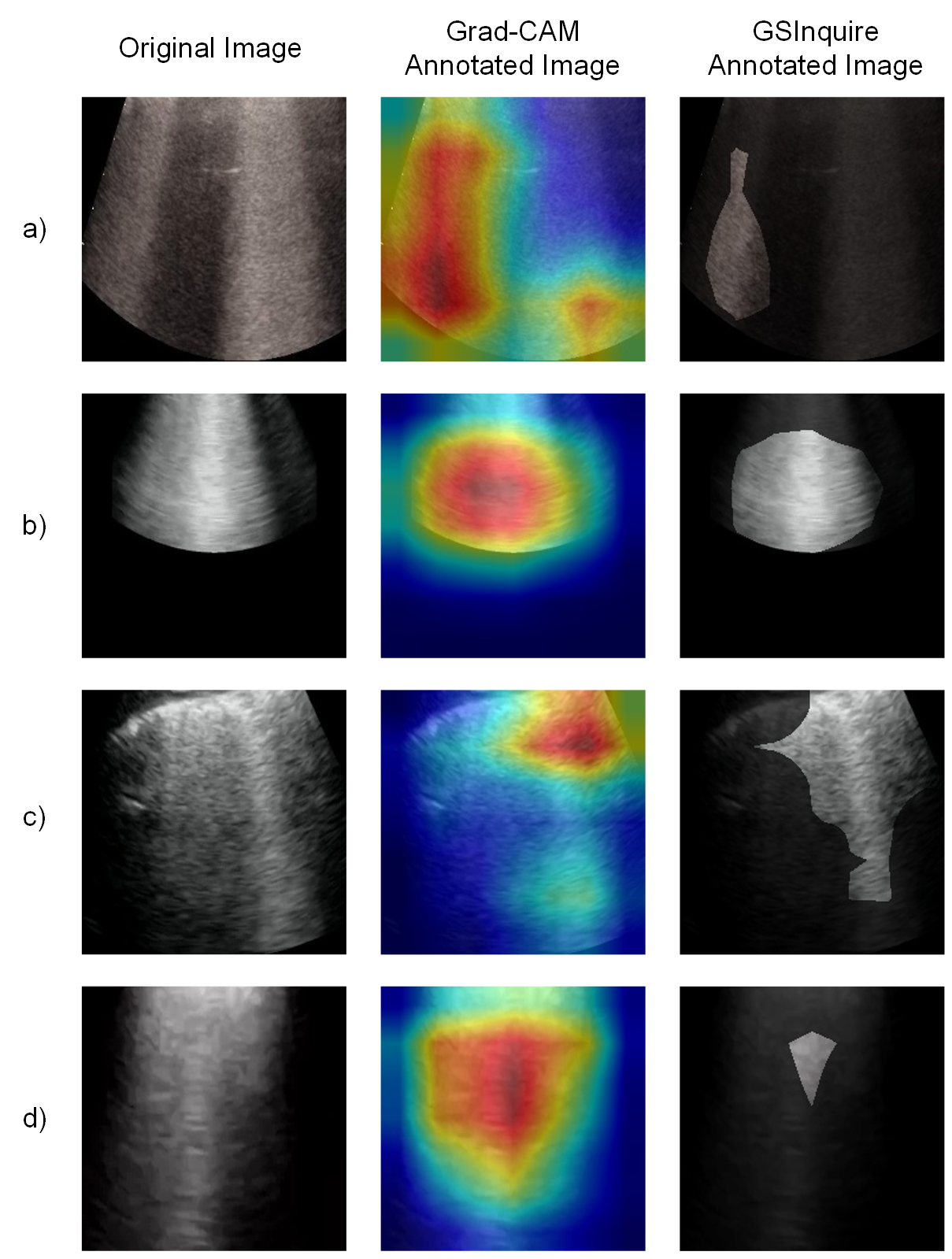}
\caption{\label{fig:cropped}Four cropped COVID-19 positive examples predicted correctly with high confidence by COVID-Net USPro (\textbf{a-d}), while recognizing disease artifacts, e.g., extended B-lines.}
\end{figure}

We implement two strategies to solve the mentioned issues and improve classification explainability. First, excluding images with low image quality, such as insufficient image depth or the lack of representative features. A severity grade introduced by COVIDx-US dataset v1.4, called lung ultrasound score (LUSS), rates each ultrasound video on a scale of 0 to 3, where 0 corresponds to presence of only normal features, and 3 corresponds to presence of severe disease artifacts \cite{COVIDxUS2021}. Therefore, in the first attempt to improve the network, images from videos with score of 0 for the normal case and images from videos with score of 2 and 3 for COVID-19 case are used to train a binary classification version of the network. By observing the annotated images, network shows to focus more on the bottom regions of the images, though cases where network focus on the top pleura region are still present. The second strategy to further improve model explainability is to exclude regions above the pleura (i.e., soft tissue) of the images, so that network focuses on the disease-defining features, present mostly at the bottom of the images below lung pleura. Our experiments confirm the effectiveness of this strategy. Hence, combining the first and second strategy, a binary model with LUSS score filtered and cropped images is trained. Figure~\ref{fig:cropped} shows examples from the cropped images analysis. As suggested from the annotated examples and confirmed by our contributing clinician (A.F.), clinically determining artifacts such as B-lines and lung consolidation are clearly identified in COVID-19 positive images by COVID-Net USPro.

\section{Conclusions}
\label{sec:discussion}
Deep neural network architectures have shown promising results in a wide range of tasks, including predictive and diagnostic tasks. However, such networks require a massive amount of labelled data to train which is against the nature of new pandemics and novel diseases where there are no or very few data samples available, especially in the initial stages. As part of the COVID-Net initiative and using a diverse complex benchmark dataset, i.e., COVIDx-US, in this work we introduce the COVID-Net USPro network, tailored to detect COVID-19 infection with high accuracy from very few ultrasound images. The proposed deep prototypical network leverages pretrained models with tailored parameters on final layers to reduce computational complexity and achieve high classification performance when only 5 examples from each class are available for training. Accuracy, precision and recall for the best performing network are over 99\%, which are comparable or outperforming other existing work \cite{COVIDNetUSMacLean2021, AdaptiveFewShotCOVIDUS}. These properties are not only highly crucial for the control of the COVID-19 pandemic but also for screening patients in new diseases/pandemics for which the proposed network can be easily tuned. We intensively assessed the explainability of the network and clinically validated its performance. Experimental results demonstrate that COVID-Net USPro can not only achieve high performance in terms of accuracy, precision, and recall, but also shows predictive behaviour that is consistent with clinical interpretation, as validated by our contributing clinician (A.F.). In addition, as part of the explainability-driven performance validation process, we proposed and implemented two strategies to further improve the network performance in accordance with the background clinical knowledge in identifying COVID-19 positive and negative cases. 
Overall, we believe the simplicity and effectiveness of COVID-Net USPro makes it a promising tool to aid the COVID-19 screening process using ultrasound images. We hope the open-source release of COVID-Net USPro help researchers and clinical data scientists to accelerate innovations in the combat against the COVID-19 pandemic that can ultimately benefit the larger society. 

Several future research directions can be explored to further improve the network. First, some additional steps in data augmentation and preparation can be taken to improve data quality and dataset size. In this work, ultrasound images captured with linear probe are excluded due to differences in clinical interpretation of linear probe and convex probe captured images. More image augmentation and preparation techniques can be experimented to include linear probe data and increase the data size. Second, in this work, we used simple cropping to filter out the pleura region of the images. A more procedural image segmentation step could be added to include only clinically relevant areas of the images for network construction to further improve network performance from the explainability standpoint. Lastly, we used COVIDx-US which is a public dataset that includes data of various sources and quality. Network training could be improved by only using high quality input ultrasound data, collected systematically, which contain clear representative image artifacts with sufficient/specific image depth. For this purpose, a data collection protocol might be required to capture ultrasound images in a standardized manner from a set of consented participants.

\bibliographystyle{unsrt}  
\bibliography{main}  

\end{document}